\documentclass[prl,aps,reprint,showpacs,superscriptaddress,amsmath]{revtex4-1}

\usepackage{bm}
\usepackage{graphicx}
\usepackage{dsfont}
\usepackage[usenames]{color}
\usepackage{epstopdf}
\usepackage[utf8]{inputenc}

\newcommand{\ket}[1]{\, | #1 \rangle}
\newcommand{\braket}[2]{\langle #1 | #2 \rangle}

\newcommand{\om}{\omega}

\newcommand{\Ga}{\Gamma}
\newcommand{\de}{\delta}
\newcommand{\De}{\Delta}

\newcommand{\eps}{\epsilon}

\newcommand{\E}{\hat{\mathcal{E}}}
\newcommand{\al}{\alpha}
\newcommand{\p}{\hat{\sigma}_{ge}}
\usepackage{cancel,ifthen,ulem}
\newcommand{\comment}[2][NoInPuT]{\ifthenelse{\equal{#1}{NoInPuT}}{}{\strike{#1}} {\color{blue} #2}}
\newcommand{\strike}[1]{{\color{red}\sout{#1}}}

\begin{document}

\title{Interfacing Superconducting Qubits and Telecom Photons via a Rare-Earth Doped Crystal} 

\author{Christopher O'Brien}
\email{cobrien.physics@gmail.com}
\affiliation{Fachbereich Physik und Forschungszentrum OPTIMAS,
Technische Universit\"at Kaiserslautern, D-67663 Kaiserslautern, Germany}
\author{Nikolai Lauk}
\affiliation{Fachbereich Physik und Forschungszentrum OPTIMAS,
Technische Universit\"at Kaiserslautern, D-67663 Kaiserslautern, Germany}
\author{Susanne Blum}
\affiliation{Theoretische Physik, Universit\"at des Saarlandes, D-66123
Saarbr\"ucken, Germany}
\author{Giovanna Morigi}
\affiliation{Theoretische Physik, Universit\"at des Saarlandes, D-66123
Saarbr\"ucken, Germany}
\author{Michael Fleischhauer}
\affiliation{Fachbereich Physik und Forschungszentrum OPTIMAS,
Technische Universit\"at Kaiserslautern, D-67663 Kaiserslautern, Germany}

\date{\today}
\begin{abstract}
We propose a scheme to couple short single photon pulses to superconducting qubits. An optical photon is 
first absorbed into an inhomogeneously broadened  rare-earth doped crystal using controlled reversible inhomogeneous broadening. 
The optical excitation is then mapped into a spin state using a series of $\pi$-pulses and subsequently transferred to a superconducting qubit via a microwave cavity.
To overcome the intrinsic and engineered inhomogeneous broadening of the optical and spin
transitions in rare-earth doped crystals, we make use of a special transfer protocol
using staggered $\pi$-pulses. We predict total transfer efficiencies on the order of 90\%.
\end{abstract}


\maketitle
%%%%%%%%%%%%%%%
%  Introducing the superconducting qubit
%%%%%%%%%%%%%%%%
Superconducting quantum circuits are a promising candidate for scalable quantum
computation. Impressive improvements have been made in the last few years: coherence times
were increased by new qubit designs \cite{Paik_2011,Chow_2012}, single qubit
operations were performed \cite{Chow_2010}, and by coupling two qubits to a
microwave cavity \cite{Blais_2004,Wallraff_2004} the realization of two qubit
gates was shown
\cite{Blais_2007,Majer_2007,DiCarlo_2009,DiCarlo_2010,Dewes_2012,Poletto_2012}. 
One major limitation of superconducting qubits (SCQ) is that they are
essentially stationary, with both qubits and microwave cavities fixed on a
single chip. 
Telecom-wavelength photons on the other hand, are the best candidates for
transporting quantum information, due to the availability of low loss optical
fibers. By interfacing short-pulse photons and SCQs, a fast quantum network could be realized
where one uses the stationary SCQ for quantum information processing and photons
for communication between different nodes. 

\begin{figure}[t]
\centering
\includegraphics[width=8cm]{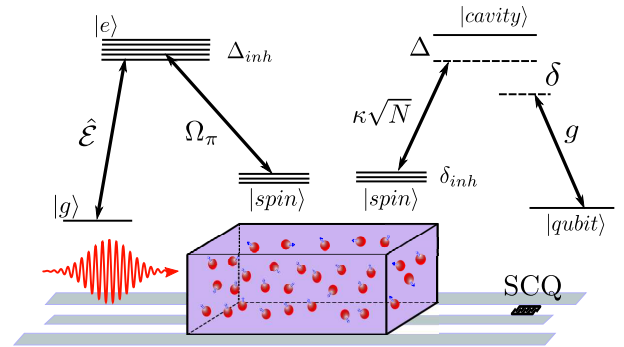}
\caption{(color online). Coherent and reversible transfer from telecom-wavelength photons
to a superconducting qubit excitation can be achieved with rare-earth doped crystals coupled to a tunable superconducting cavity. Left: A single photon is absorbed into a collective spin excitation in the REDC. Right: The crystal spin is transferred into an
excitation of the SCQ.}
\label{fig:scheme}
\end{figure}

There are several proposals for such hybrid devices \cite{Hybrid, Stannigel_2010}. One direction is the use of nanomechanical devices \cite{Stannigel_2010, Barzanjeh_2014}, which strongly couple to both microwave and optical fields \cite{Bachmann_2013,Andrews_2014,Bagci_2014}.
Another direction is to couple a  microwave cavity with 
a spin ensemble, which could be a cold gas \cite{Imamoglu_2009, Verdu_2009}, NV-centers in diamond \cite{Kubo_2010}, or rare-earth doped crystals \cite{Staudt_2012,Probst_2013}. Coupling spin states of NV-centers to a SCQ via a microwave cavity has been realized recently \cite{Kubo_2011}, also direct coupling of the 
spin ensemble to a superconducting flux qubit was proposed and implemented in \cite{Marcos_2010,Zhu_2011}.
However, NV-centers have the disadvantage of being
incompatible with the telecom bandwidth.
In principle wavelength incompatibility could be addressed by single photon frequency conversion \cite{Zaske2012,Blum2013}.  This
issue can be overcome by using rare-earth doped crystals.
Inspired by the recent experimental demonstrations of coupling a spin ensemble
in a rare-earth doped crystals (REDC) to a microwave cavity \cite{Staudt_2012,Probst_2013,Bushev_2014} we
here present and analyze a scheme for coupling optical photons to
superconducting qubits mediated by a REDC (see Fig.\ref{fig:scheme}).
The large inhomogeneous broadening of optical and spin transitions in 
REDCs allows for an interface between SCQs and short photon pulses and thus 
paves the way for a high bit-rate quantum network.
%%%%%%%%%%%%%%%%%%%%%%%%%%%%%%%%%%%%%%%%%%%%%%%%%%%%%%%%%%%%%%%%%%%%%%%%%%%%%%%
%%%%%%%%%%%%%%%%%%%%%%%%%%%%%%%%%%%%%%%%%%%%%%%%%%%%%%%%%%%%%%%%%%%%%%%%%%%%%%%

However, such a large broadening of the optical transition, as well as the associated broadening of the spin-transition leads to fast dephasing of the symmetric collective state and therefore a very different approach than that used in NV-centers is required. This broadening, as well as the associated broadening of the spin-transition, on the other hand leads to fast population transfer from the symmetric collective state into non-symmetric ones, which do not couple to the microwave resonator. Here we propose and theoretically analyze a specific protocol using a staggered series of $\pi$-pulses which compensates the effect of the  induced broadening and facilitates the transfer to a SCQ with above 90\% efficiency.

The total efficiency for transfer between the telecom photon and a SCQ can be
expressed as:
$\eta = \eta_{S} \eta_{T}$, where $\eta_{S}$ is the efficiency of storing a
photon as a collective spin state that can be directly coupled to the
microwave cavity, and $\eta_{T}$ is the efficiency of moving the collective
spin-excitation to an excitation of the qubit. We will see that $\eta_S$ is 
essentially determined by the spectral shape of the input pulse and $\eta_T$ is
dominated by the incoherent loss of population during the transfer process to non-coupled spin states due to
intrinsic inhomogeneous broadening of the spin state.
Whereas the dephasing due to 
\textit{induced} broadening can be reversed, population losses due to \textit{intrinsic} broadening 
constitute the main bottleneck of any transfer protocol to microwave photons. 

We first analyze the efficiency $\eta_S$ for storing a photon in a collective spin state. 
We suggest following one of the possible implementations of the CRIB protocol where the transition
frequency is a function of distance into the medium in the direction of photon propagation
\cite{Hetet_2008,Sparkes_2013}. The preparation protocol is as follows: Starting with the
inhomogeneously broadened optical line, hole burning is performed by optically pumping
all of the atoms into a long-lived shelving state. Then using a narrow band laser, some of the atoms
are pumped back into the excited state such that there is now a narrow frequency
peak. The remaining atoms in the shelving state do not participate in the optical interaction. Finally this peak is frequency broadened by applying a magnetic field (Zeeman shift) linearly varying in space along the crystal length $L$. This leads to a frequency profile with slope ${\rm d}\omega/{\rm d}z =\alpha$ in the frequency interval $[\omega_0-\alpha L/2,\omega_0+\alpha L/2]$, chosen to match the frequency width of the incoming photon. As will be seen later it is important that this magnetic field gradient also
induces inhomogeneous broadening of the spin energy levels.

The efficiency $\eta_{S}$ is determined by the spatial overlap between the wave function $\ket{\Psi(t)}$ with the state $\ket{\Psi_1}=\ket{\Psi_S}\ket{vac}$, where $\ket{vac}$ is the field vacuum state and $\ket{\Psi_S}$ is the collective spin state which interacts with the cavity, thus $\eta _{S} = |\braket{\Psi_1}{\Psi(t)}|^2$.  
In turn, $\ket{\Psi_S}$ is determined by the cavity mode function, and thus by the structure of a coplanar waveguide resonator mode \cite{Schuster_2010, Verdu_2009}. In order to fulfill the spatial mode matching condition, one can perform hole burning not only in the frequency domain but also spatially. By doing this, one can create an interaction region which is small enough transversely, such that the mode function could be assumed to be constant on that length. The interacting spin state $\ket{\Psi_S}$ is then approximated by the symmetric Dicke state $\ket{\Psi_S} =\sum_j\ket{1_j}_s/\sqrt{N}$, where $\ket{1_j}_s=\ket{g}_1\ket{g}_2\ldots\ket{s}_j\ldots\ket{g}_N$ represents the state with the $j$-th atom in the spin state $\ket{s}$ and the rest in the ground state $\ket{g}$. The sum runs over the $N$ crystal spins that are in the interaction region of
the cavity.

We assume that the initial state is $\ket{\Psi(0)}=\ket{\{0\}_s}\ket{1_f}$, where $\ket{\{0\}_s}$ describes all atoms in $\ket{g}$ and $\ket{1_f}$ denotes the state of a incident single photon. The wave function $\ket{\Psi(t)}$ at $t>0$ is found after a two-step process. First, the photon is absorbed into the optical transition $\ket{e}-\ket{g}$. We can describe this process by linearized equations of motion for the slowly varying operators for the probe field $\E$ and the optical coherence $\p$ with spatial dependent detuning in the co-moving frame:
\begin{align}
  \partial_t \p(z,t)&=-i \al z \p(z,t)+i g_{eg}\E(z,t), \\
  \partial_z \E(z,t)&=i\frac{g_{eg} n}{c}\p(z,t),
\end{align}
where $g_{eg}=d_{ge}\sqrt{\frac{\om_s}{2 \eps_0\hbar A}}$ is the coupling
constant with optical resonance frequency $\om_s$, beam cross-section $A$, atomic density $n$, and transition dipole moment $d_{ge}$. We neglect spontaneous emission from
the excited state, since for a REDC the decay time is much longer than the usual pulse times. 

To ensure that the incoming photon is fully absorbed we require the photons spectral
width $\De\om_p$ to fit within the induced inhomogeneous broadening such that,
 $\De\om_p < \al L$. In this limit we follow the analysis of \cite{Moiseev_2008} and approximate the solution for the coherence by:
\begin{align}
 &\p(z,t)=-g_{eg} e^{-\frac{\pi d}{2}}\int_0^t dt'\frac{e^{-id\log[(t-t')\al L/2]}}{d\Ga(-id)}
 \nonumber \\& 
\cdot e^{-i\al z(t-t')}\E(-L/2,t'),
\end{align}
where $\Ga(ix)$ is the complex gamma function ($x$ real) and $d=g_{eg}^2n/c\al$ is the
optical depth of the medium. For $t$ much larger than the pulse time,
$t \gg T_p$, this expression can be cast into the form:
\begin{align}
 \p(z,t)=-g_{eg} e^{-\frac{\pi d}{2}}\frac{e^{-id\log(t\al L/2)}}{d\Ga(-id)} e^{-i\al zt}\sqrt{2\pi}\tilde{\mathcal{E}}(-L/2,\al z) \label{eq:sol}
\end{align}
where $\tilde{\mathcal{E}}(-L/2,\om) = \frac{1}{\sqrt{2\pi}}\int dt e^{i \om t} \E(-L/2,t)$ is the Fourier transform of the incoming field at the
beginning of the medium. The phase term $\exp(-i\alpha zt)$ is due to the induced inhomogeneous
broadening, which is essential to prevent reemission of the excitation as a photon echo.  After the photon is absorbed
into the optical transition $\ket{e}-\ket{g}$ the corresponding coherence starts to dephase. The induced inhomogeneous broadening can be controlled in order to rephase the excitation at a specific time. 
To that end, in the second step we use a series of three $\pi$-pulses between the spin and excited
levels. At a time $\tau_1$ a $\pi$-pulse on the $\ket{e}-\ket{s}$ transition is applied to move the population
into the spin state, where it continues to dephase due to the inhomogeneous broadening of the spin state, but now with a different rate. 
Then after some time $\tau_2$ we simultaneously flip the magnetic field gradient \footnote{The switching times of the magnetic fields have to be sufficiently small compared to the dephasing time, which is challenging but within reach of current technology.} to begin rephasing while using a $\pi$-pulse to
bring the population back into $\ket{e}$. Once the atoms have spent the same time
in the excited state as during storage, the dephasing due to inhomogeneous broadening of the excited state is compensated and the $\exp(-i\alpha z t)$ term vanishes. Note however,
that due to the accumulated phase caused by the inhomogeneous broadening of the spin state there will be no photon echo.
Then applying the third $\pi$-pulse we again move the excitation to the spin state, where it can now complete rephasing.
For faithful transfer the pulses must have a Rabi frequency larger then the
inhomogeneous width of the spin transition, and be well timed to get the proper pulse area.

Under these assumptions the evaluation of $\eta_S$ reduces to $\eta_S= |\braket{\{0\}_s,vac}{(\sum_k\sigma_{ge}^k(t)/\sqrt{N})|\{0\}_s,1_f}|^2$,
and in the continuum limit, using Eq.(\ref{eq:sol}) we find
\begin{align}
\eta_{S}=|\braket{\Psi_1}{\Psi(t)}|^2 = (1-e^{-2\pi d})\left|\int_{-1/2}^{1/2}d \xi \bar{\mathcal{E}}(\xi)\right|^2\,, \label{eq:overlap}
\end{align}
where $z=\xi L$, $\tau=\al L t $, and $\bar{\mathcal{E}}=\tilde{\mathcal{E}}\sqrt{L\alpha c}$ is the dimensionless field envelope. For optical depths $d>1$ the overlap is thus given by the square of the integral of the input field's Fourier transform which is maximal for constant $\bar{\mathcal{E}}$, whereas
a photon with Gaussian envelope could reach $\eta_S>90\%$. 

As a next step, we discuss the transfer dynamics of the spin excitation to a SCQ. 
The combined spin-cavity-qubit dynamics, sketched on the right of Fig.\ref{fig:scheme}, is governed by the
Hamiltonian:
\begin{align}
&H = \sum_j \hbar \omega _j \hat S^z _j + \sum _j \hbar \kappa \left(\hat a^\dagger \hat S_j + \hat S_j ^\dagger \hat a \right) 
\nonumber \\& 
+ \hbar \omega_c \hat a^\dagger \hat a + \hbar G \left( \hat \sigma ^\dagger \hat a + \hat a^\dagger \hat \sigma \right) + \hbar \omega_q \hat \sigma ^z,
\end{align}
where $G$ is the coupling between the cavity and the qubit. $\hat \sigma$ is the qubit
spin flip operator, and $\hat \sigma ^z$ describes spin
population operator for the SCQ. $\kappa$ denotes the coupling between an individual spin in
the REDC and the cavity and is assumed to be the same for every interacting spin.
$\hat a$ and $\hat a^\dagger$ are the annihilation and creation operators
of the microwave cavity mode. $\hat S_j$ is the collective spin flip operator in the REDC, $\hat S^z_j$ corresponds spin population operator,
$\omega _j$ is the frequency of each spin, $\omega _c$ is the resonance frequency for the cavity, and $\omega_q$ is the frequency of the SCQ.

To discuss the transfer protocol let us consider a quantum state with a single excitation for $t'>t$:
\begin{align}
&\ket{\Psi(t')} = \sum_j \xi_j (t') \ket{1_j}_s \ket{0}_c \ket{\downarrow}_q 
\nonumber \\ &
+ c(t') \ket{ \{0 \}}_s \ket{ 1}_c  \ket{\downarrow}_q + q(t') \ket{\{0\}}_s\ket{0}_c \ket{\uparrow}_q, \label{eq:singleexc}
\end{align}
where $\ket{0}_c$ and $\ket{1}_c$ are zero and one photon states of the cavity mode, while $\ket{\downarrow}_q$ and $\ket{\uparrow}_q$ are the qubit ground and excited states.  The initial probability amplitudes are such that $\xi_j(t_0)=1/\sqrt{N}$ and $q(t_0)=c(t_0)=0$. In order to efficiently transfer the population between spin and qubit states one could resort to STIRAP like procedures, but these are precluded as they would require one to change the coupling constants $G$ and $\kappa$ as a
function of time. However, one can control the resonance frequency of the cavity and of
the spin states by adjusting the applied fields. Therefore adiabatic transfer of population could be implemented by sweeping the cavity frequency through resonance first with the spin state and then with the qubit.
Since adiabaticity requires that the population spends a long time in the spin state, which has large inhomogeneous broadening, this adiabatic scheme is too slow to have faithful transfer in available REDC.

\begin{figure}[t]
\centering
\includegraphics[width=8.5cm]{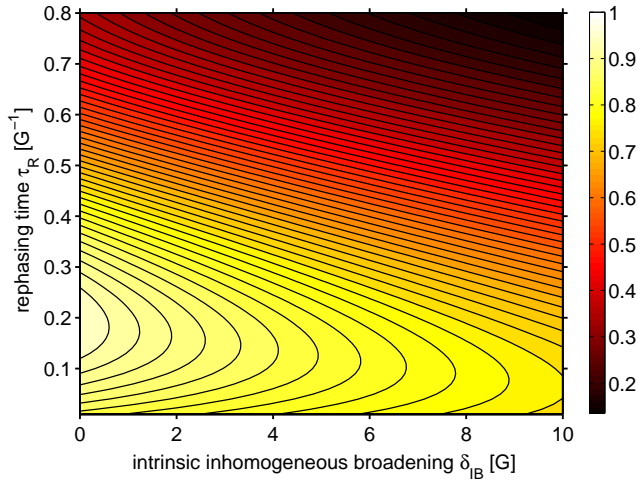}
\caption{
(color online). Transfer efficiency as a function of the intrinsic inhomogeneous broadening $\delta_{IB}$ and the rephasing time $\tau_R$, for a fixed induced broadening of $\delta_{inh} = 7G$
and $\kappa\sqrt{N} = 6G$.}
\label{fig:staggered}
\end{figure}

We nevertheless identified an alternative protocol which is based on a series of staggered
$\pi$-pulse interactions and warrants efficient population transfer.
The protocol requires that qubit and spin state are initially far detuned from each other. The cavity is then brought into
resonance with the spin for an amount of time $T_S$ corresponding to a $\pi$-pulse,
$T_S = \pi/(2\kappa \sqrt{N})$, which transfers the population into one cavity excitation. Then the cavity frequency is tuned into resonance with the SC-qubit for a time $T_C$ corresponding to another $\pi$-pulse, $T_C = \pi/(2G)$. Finally,
the cavity is detuned once again, leaving the excitation in the qubit. To simulate this process we numerically solve the Schr\"odinger equation for the coefficients in Eq. (\ref{eq:singleexc})
with time dependent detunings, whereby we disregard the cavity decay since the stripline cavities have typically high quality factors. On first glance one might think that the induced inhomogeneous broadening $\de_{inh}$ must be kept much smaller than the  REDC-cavity coupling $\kappa\sqrt{N}$
to avoid dephasing during $T_S$, which would substantially limit the bandwidth of the transfer. The results of our simulation show however that by timing the transfer procedure to match up with the rephasing of the spin excitation
the transfer protocol can be quite efficient, even with high inhomogeneous broadening. This is demonstrated in Fig.\ref{fig:staggered} where we plot the transfer efficiency as a function of the intrinsic inhomogeneous broadening
and the rephasing time $\tau_R$ when the protocol is started. 

We note that by reversing our protocol the qubit excitation 
can be transferred back into a superposition state of a single photon and vacuum. This includes the hole burning steps which are important to ensure the collective spin state is retrieved as a photon wave packet.
%
%%%%%%%%%%%%%%%%%%%%%%%%%%%%%%%%%%%%%%%%%%%%%%%%%%%%%%%%%%%%%%%%%%%%%%%%
%%%%%%%%%%%%%%%%%%%%%%%%%%%%%%%%%%%%%%%%%%%%%%%%%%%%%%%%%%%%%%%%%%%%%%%
%%%%%%%%%%%%%%%%%%%%%%%%%%%%%%%%%%%%%%%%%%%%%%%%%%%%%%%%%%%%%%%%%%%%%%%%
\begin{figure}[t]
\centering
\includegraphics[width=7cm]{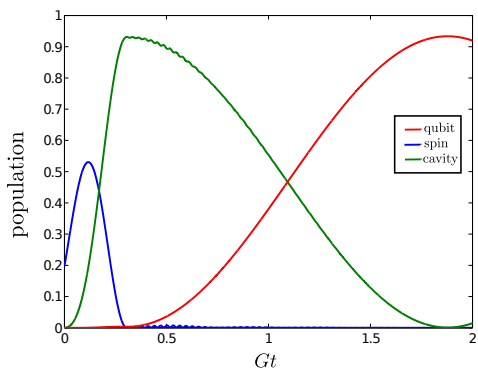}
\caption{(color online). Numerical modeling of staggered transfer showing a 92\% transfer, with an initial cavity detuning $\Delta = 20G$ and with $\kappa \sqrt{N} = 6G$.
With an intrinsic broadening of $\delta_{IB} = 2G$ and an induced broadening of $\delta_{inh} = 10G$,
which is completely compensated through rephasing of the symmetric Dicke state with an rephasing time of $G\tau _R = 0.15$.}
\label{fig:ErYSO}
\end{figure}
%%%%%%%%%%%%%%%%%%%%%%%%%%%%%%%%%%%%%%%%%%%%%%%%%%%%%%%%%%%%%%%%%%%%%%%
%

With proper pulse timings the loss of efficiency is only due to the intrinsic
broadening $\de_{IB}$ of the spin state causing population loss to the uncoupled spin modes.
To estimate these losses we can make following considerations: If all of the spins are in resonance, then only the symmetric Dicke state interacts with the cavity. This is the state that would be created by our storage process, if there is no intrinsic inhomogeneous broadening. The presence of this intrinsic broadening couples the Dicke state to the other $N-1$ non-symmetric eigenmodes, so the actual state collective spin state take the form $\ket{\Psi_{inh}(t)} = \sum _j e^{i\delta_j t}\ket{1_j}_s/\sqrt{N} $, with the sum over intrinsic spin frequency detunings $\delta_j$.
Using this expression we estimate the influence of the intrinsic broadening in our protocol
by $\eta_T=|\braket{\Psi _S}{\Psi_{inh}(T_S)}|^2$. Assuming a homogeneous excitation profile it takes the form
\begin{align}
 \eta_{T} = \frac{\sin(\delta_{IB} T_S)^2}{(\delta_{IB}T_S)^2} \approx e^{-\frac{1}{3} (\delta_{IB} T_S)^2}\,.
 \label{eq:flat}
 \end{align}
This dephasing is thus a loss mechanism in the transfer process, which limits the maximum time $T_S$ that can be spent in the spin state.

While we considered only the transfer of a single photon excitation, the vacuum state will also be faithfully preserved since there are no stray photons added to the strip line cavity. Therefore, the efficiency of transferring an arbitrary initial photon state $\ket{\phi_f}$ is given by the efficiency of transferring a single photon excitation. Assuming all dynamical phases are well under control, i.e. the storage process preserves coherence 
then the  intrinsic inhomogeneous broadening $\de_{IB}$ is the only dephasing source.

For the experimental realization of our scheme we propose to use $\text{Er}^{3+}\text{:YSO}$ for which CRIB has been demonstrated \cite{Gisin_2010}. It has been shown \cite{Probst_2013}, that a collective spin-cavity coupling of $\kappa \sqrt{N} = 34$MHz can be achieved with the HWHM of the intrinsic inhomogeneous
broadening being reducible to $\delta_{IB} = 12$ MHz. Note that spectral hole burning will reduce the effective number of interacting spins $N$.
Although, the SCQ is acting as a single spin, it has a very
strong dipole moment \cite{Blais_2004}, and thus strong coupling between the qubit and cavity is possible with coupling
strength $G$ of several MHz. Making an estimate for an implementation in Er:YSO with the given numbers,
the best transfer efficiency can exceed $\eta  \geq 90\%$, as shown in Fig.\ref{fig:ErYSO}. This transfer efficiency is mainly limited by the achievable mode overlap which may improve with further tuning, by the intrinsic inhomogeneous broadening of both the spin and excited states in rare-earth doped crystals, and by the need for spatial hole burning which reduces the number of interacting ions and requires a suitable shelving state. Better crystals, or better dopants may be grown in the future with larger dipole moments, higher concentrations, less broadening, and long lived shelving states, that will allow the efficiency to be higher.

A draw back of the proposed staggered protocol is the necessity to be able to tune the cavity over a large range. To get perfect transfer it may be required to tune the $6-10$ GHz cavity up to a $1$ GHz in frequency. Some of the reported
experimental systems allow tuning \cite{Osborn_2007} this large but the average seems to be below $500$ MHz \cite{Sandberg_2008}. One could use less sweeping at the cost of the transfer efficiency. Alternatively, one can take advantage of $G$ being weak in comparison to the collective spin coupling, by starting at a smaller fixed cavity-qubit detuning $\delta$, and simultaneously detuning the spin state far away from resonance when the cavity is tuned into resonance with the qubit.
In this case there is a factor of 6 improvement in minimizing the  cavity sweeping at the cost of needing to strongly detune the spin which should be experimentally easier, however. This would bring the required change in the cavity resonance to less than $200$ MHz, which is currently achievable. 

In summary we analyzed the efficiency of a scheme to faithfully and reversibly transfer a quantum state 
between short optical photons and a superconducting qubit. It relies on coupling rare-earth doped
crystals with engineered inhomogeneous broadening to a microwave cavity, that in turn couples to a superconducting qubit. The incoming photon needs to be stored in the spin coherence as a symmetric Dicke state which can be achieved using
the controlled reversible inhomogeneous broadening quantum memory protocol with a spatially varying frequency. The main loss mechanism for the transfer procedure is the intrinsic inhomogeneous broadening of the spin state. To overcome this it is necessary to limit the time that the pulse remains as a coherent spin excitation. The best way to accomplish this is to forgo adiabatic transfer schemes, and to use a  staggered $\pi$-pulse transfer scheme with proper timing adjusted such that the engineered broadening is compensated. Increasing the effective spin-cavity coupling will give another improvement by reducing the transfer time. Combining proposed storage and transfer schemes, could lead to successful transfer of a short single photon pulse to a SCQ using a Er:YSO crystal with transfer efficiencies on the order of $90\%$.
\begin{acknowledgments}
The authors would like to acknowledge useful discussions with Pavel Bushev, and 
financial support by the German Federal Ministry of Education and Research
(BMBF, project QuOReP 01BQ1005 and 16BQ1011) and by the German Research Foundation (DFG).
\end{acknowledgments}

\bibliographystyle{apsrev4-1}
\bibliography{Opt_Micro_Paper_resubmit}

\end{document}